# An Overview of the New Test Stand for H- Ion Sources at FNAL[a]

A. Sosa,[1,b] D. S. Bollinger,[1] K. Duel,[1] P. Karns,[1] W. Pellico[1] and C. Y. Tan[1]

[1]Fermi National Accelerator Laboratory, P.O. Box 500, Batavia, IL 60510-5011, USA

A new test stand at FNAL is being constructed to carry out experiments to develop and upgrade the present magnetron-type sources of H- ions of up to 80 mA at 35 keV in the context of the Proton Improvement Plan (PIP). The aim of this plan is to provide high-power proton beams for the experiments at FNAL. The technical details of the construction and layout of this test stand are presented, along with a prospective set of diagnostics to monitor the sources.

## I. MAGNETRON H- SOURCES AT FNAL

### A. Introduction

The H- injector at Fermi National Accelerator Laboratory (FNAL) operates continuously for up to a year or more. Requested uptime is 90% with only four weeks of planned downtime. The intensive use of the H- dimpled magnetron sources currently used at FNAL severely limits the capacity of carrying out tests on the operational sources. This situation motivates the need for a dedicated, off-line test stand where a new R&D effort is underway to develop these sources and improve their reliability and performance.

### B. Source Parameters

The H- magnetron ion source installed in the test stand is an exact replica of the style of source used in the operational machine. A model of this source is depicted in Fig. 1. This cesiated source delivers H- beam at an energy of 35 keV with a nominal beam current between 60 to 80 mA. The source and its body are attached to a metallic structure which is biased to -35 kV and pulsed at 15 Hz. The source has a circular aperture of 3.175 mm (0.125") and operates under a vacuum level of ~$10^{-6}$ Torr. This source uses surface production of H- ions from a cesium-coated molybdenum cathode. $H_2$ gas is injected into the source using a valve pulsed at 15 Hz. A plasma is sustained with an arc voltage of ~200 V between two electrodes and a permanent magnet provides the perpendicular magnetic field. In this E×B region, protons are attracted to the cathode's surface for H- production, and electrons flow around the race-track shaped volume between anode and cathode, as shown in Fig. 1. The key source parameters are listed on Table 1.

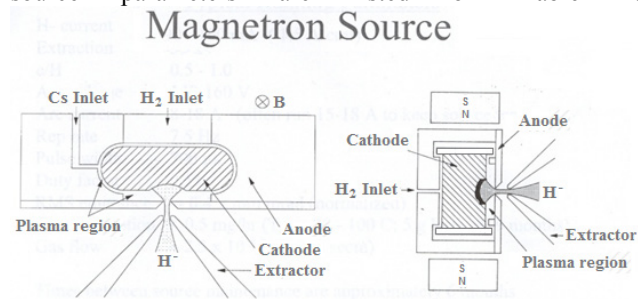

FIG. 1. Schematic of the magnetron H- source.

TABLE I. FNAL H- Source Parameters.

| Parameter | Value | Units |
|---|---|---|
| Arc Current | 15 | A |
| Arc Voltage | 180 | V |
| Extractor Voltage | 35 | kV |
| Beam Current | 80 | mA |
| Power Efficiency | 48 | mA/kW |
| Rep Rate | 15 | Hz |
| Arc Pulse Width | 250 | µs |
| Extracted Beam Pulse Width | 80 | µs |
| Duty Factor | 0.375 | % |
| Cathode Temperature | 380 | °C |
| Cs Boiler Temperature | 130 | °C |
| Emittance $\varepsilon x/\varepsilon y$ | 0.17/0.28 | π mm mrad |
| Extraction Gap | 4.67 | mm |
| Lifetime | 9 | months |

## II. TEST STAND FOR H- SOURCES

The test stand, pictured in Fig. 2, provides an off-line location for machine development studies.

---

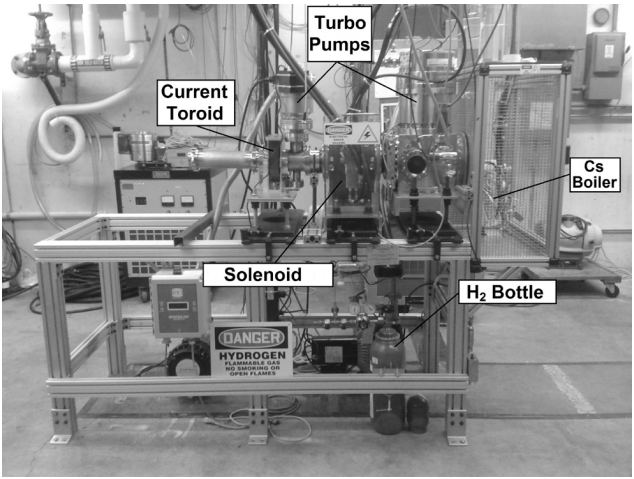

FIG. 2. Test stand for negative H sources.

The mechanical support is made of aluminum, which is grounded by means of welded copper sheets on the floor. The vacuum system is comprised of two roughing pumps and two turbo pumps. An Edwards STPA1603C turbo-molecular pump connected to the source vacuum chamber shown in Fig. 3, provides a pumping capacity of 1,200 l/s for $H_2$. The other turbo-molecular pump, which is at the end of the Low Energy Beam Transport line (LEBT), is an Oerlikon-Leybold TMP361 with a pumping capacity of 340 l/s for $H_2$. The source vacuum chamber is made of stainless steel with 10 inch CF flanges on all sides. The high-voltage ceramic insulator allows the source to be biased at -35 kV, whereas the extractor cone is connected to ground through the source vacuum chamber by means of metallic connectors on the downstream side.

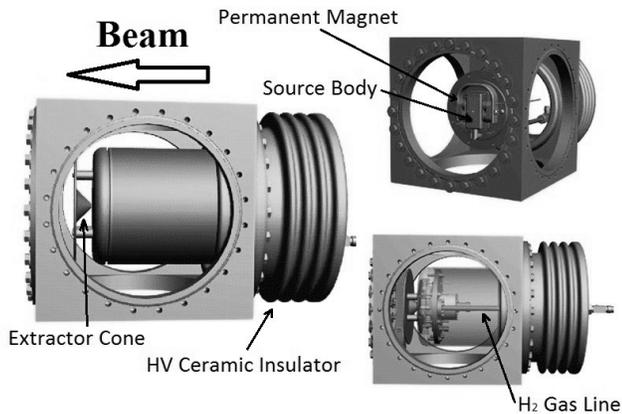

FIG. 3. 3D model of the source and its vacuum chamber.

## III. ELECTRONICS

The following set of electronics is necessary to operate and monitor the source. First, the gas valve power supply and its pulser, with a maximum voltage of 300 V. These allow $H_2$ gas to be pumped into the source at 15 Hz where the gas pressure changes according to the gas valve voltage. Then the arc power supply and its pulser, which deliver a voltage of up to -300 V between anode and cathode necessary to maintain an arc current in the source. The arc is typically 1 ms after the end of the gas pulse. The extractor power supply and its pulser are able to deliver 35 kV while pulsing the whole source assembly at 15 Hz. In addition, a set of power supplies and their temperature reading modules for the different heaters (Cs boiler, valve, tube and source) are used to deliver Cs to the source. A VME crate, Smart Rack Monitor and Internet Rack Monitor complete the set of electronics that connect the source equipment to the control system.

## IV. FUTURE DEVELOPMENTS

### A. Solenoid Gas Valves

The gas valves currently used in the operational sources are piezoelectric valves. These valves flex proportionally to the applied voltage, however, they are very susceptible to subtle ambient temperature/pressure changes and they often need to be recalibrated. For this reason, a new type of valve is being tested on the test stand. It consists of a solenoid which actuates a poppet, as shown in Fig. 4. Preliminary runs allowed the source to operate in similar conditions as when using piezoelectric valves. The typical gas pulse width is about 90 μs when using a piezoelectric valve, and 200 μs when using the solenoid gas valve. More studies are being performed with this type of valve in order to assess their performance.

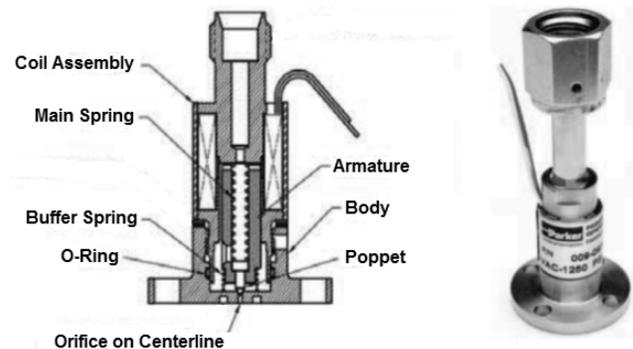

FIG. 4. Solenoid valve used for tests with H- source.

### B. Gas Mixing Experiment

One of the main experiments to be run at the test stand consists of mixing $H_2$ and $N_2$ gases at different ratios on the line that feeds the source. The idea is to verify the effect of $N_2$ on the beam current noise, as previously suggested [1]. The experimental setup is depicted in Fig. 5. Two mass flow meters calibrated to each pure gas will regulate the amount of each gas flowing into the source. These mass flow controllers measure pressure drop in laminar flows of gas (10 to 15 psi) which make them accurate down to 0.8% of the reading and 0.2% of their full scale [2]. With this setup, $N_2$ ratios between 0.5 to 3% will be tested, and the mean beam current noise will be measured connecting the output of the current toroid to an oscilloscope.

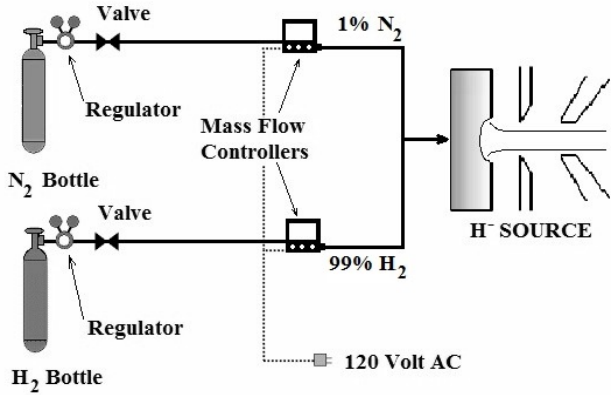

FIG. 5. Experimental setup for gas mixing into the source.

## C. New Cesium Boiler Design

The Cesium boilers currently used in the operational sources consist of a copper container where a glass ampoule with 5 g of Cs is introduced and sealed to a valve. Once the boiler is under vacuum, the boiler is squeezed, breaking the glass ampoule and releasing the Cs necessary to coat the source. In order to facilitate the handling of the Cs boilers and improve the delivery of Cs into the source, a smaller Cs boiler has been designed, shown in Fig. 6. It is made of stainless steel, with heating wire tack-welded around it to improve heat flow. Eliminating the glass ampoule introduces new challenges in order to transfer the Cs from the ampoule to the new boiler in a safe environment.

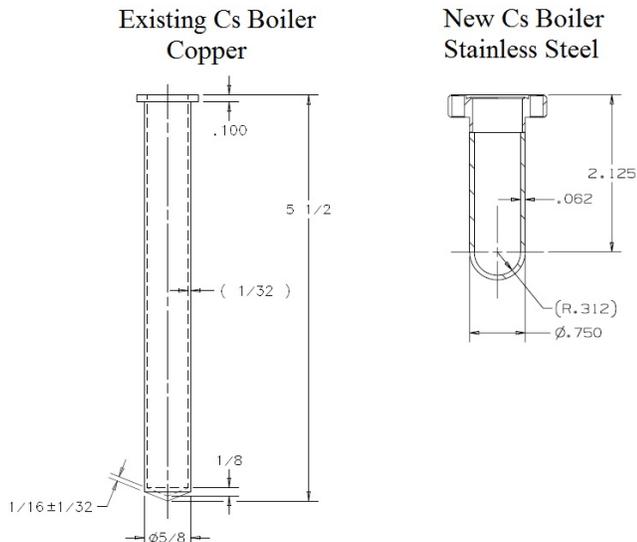

FIG. 6. Existing and new cesium boilers used in the magnetron H- source.

## D. Diagnostics

Future additions to the LEBT in the test stand include a diagnostic box, where a Faraday cup and two slit+grid emittance probes will be installed. This will allow to measure transverse beam emittance in both planes with a well-focused beam thanks to the focusing solenoid, not previously available in the older test stand. The addition of a Faraday cup will make possible to compare beam intensity measurements with respect to the current toroid.

## E. Mobile Extractor Cone

In the current design, the extractor cone is aligned and secured to the rest of the source structure with three standoffs made of Macor® [3]. Over time, contamination from the source builds up in the standoffs' surfaces that face the extraction gap leading to increased sparks. In order to reduce sparking, a modification of the design will be implemented, which will attach the extractor cone to the downstream wall of the source vacuum chamber, eliminating the need for standoffs. In addition, the extractor cone will be installed in a mobile actuator, which will make the extraction gap a new parameter to tune the source. This introduces new alignment issues whenever a source assembly is installed/removed from the source vacuum chamber that need to be addressed.

## V. CONCLUSIONS

The new test stand for H- ion sources at FNAL provides great flexibility for installation/removal of the sources and represents a step forward in the research and development effort on the magnetron H- sources currently used in the laboratory. Experimental tests will be performed in a number of issues affecting the performance of the source, specifically to improve the delivery of Cs to the source, reduce the beam current noise, and maintain a stable arc current.

## VI. ACKNOWLEDGEMENTS